\documentstyle[12pt]{article}

\catcode`\@=11
\long\def\@makefntext#1{
\protect\noindent \hbox to 3.2pt {\hskip-.9pt
$^{{\ninerm\@thefnmark}}$\hfil}#1\hfill}		

\def\@makefnmark{\hbox to 0pt{$^{\@thefnmark}$\hss}}  

\def\ps@myheadings{\let\@mkboth\@gobbletwo
\def\@oddhead{\hbox{}
\rightmark\hfil\ninerm\thepage}
\def\@oddfoot{}\def\@evenhead{\ninerm\thepage\hfil
\leftmark\hbox{}}\def\@evenfoot{}
\def\sectionmark##1{}\def\subsectionmark##1{}}

\renewcommand{\thefootnote}{\fnsymbol{footnote}}

\newcounter{sectionc}\newcounter{subsectionc}\newcounter{subsubsectionc}
\renewcommand{\section}[1] {\vspace*{0.6cm}\addtocounter{sectionc}{1}
\setcounter{subsectionc}{0}\setcounter{subsubsectionc}{0}\noindent
	{\normalsize\bf\thesectionc. #1}\par\vspace*{0.4cm}}
\renewcommand{\subsection}[1] {\vspace*{0.6cm}\addtocounter{subsectionc}{1}
	\setcounter{subsubsectionc}{0}\noindent
	{\normalsize\it\thesectionc.\thesubsectionc. #1}\par\vspace*{0.4cm}}
\renewcommand{\subsubsection}[1]
{\vspace*{0.6cm}\addtocounter{subsubsectionc}{1}
	\noindent {\normalsize\rm\thesectionc.\thesubsectionc.\thesubsubsectionc.
	#1}\par\vspace*{0.4cm}}

\newcounter{appendixc}
\newcounter{subappendixc}[appendixc]
\newcounter{subsubappendixc}[subappendixc]

\renewcommand{\appendix}[1] {\vspace*{0.6cm}
        \refstepcounter{appendixc}
        \setcounter{figure}{0}
        \setcounter{table}{0}
        \setcounter{equation}{0}
        \renewcommand{\thefigure}{\Alph{appendixc}.\arabic{figure}}
        \renewcommand{\thetable}{\Alph{appendixc}.\arabic{table}}
        \renewcommand{\theappendixc}{\Alph{appendixc}}
        \renewcommand{\theequation}{\Alph{appendixc}.\arabic{equation}}
        \noindent{\bf Appendix \theappendixc #1}\par\vspace*{0.4cm}}

\def\abstracts#1{{

\centering{\begin{minipage}{12.2truecm}\footnotesize\baselineskip=12pt\noindent
	\centerline{\footnotesize ABSTRACT}\vspace*{0.3cm}
	\parindent=0pt #1
	\end{minipage}}\par}}

\renewenvironment{thebibliography}[1]
	{\begin{list}{\arabic{enumi}.}
	{\usecounter{enumi}\setlength{\parsep}{0pt}
\setlength{\leftmargin 1.25cm}{\rightmargin 0pt}
	 \setlength{\itemsep}{0pt} \settowidth
	{\labelwidth}{#1.}\sloppy}}{\end{list}}

\newcounter{itemlistc}
\newcounter{romanlistc}
\newcounter{alphlistc}
\newcounter{arabiclistc}

\newcommand{\fcaption}[1]{
        \refstepcounter{figure}
        \setbox\@tempboxa = \hbox{\footnotesize Fig.~\thefigure. #1}
        \ifdim \wd\@tempboxa > 6in
           {\begin{center}
        \parbox{6in}{\footnotesize\baselineskip=12pt Fig.~\thefigure. #1}
            \end{center}}
        \else
             {\begin{center}
             {\footnotesize Fig.~\thefigure. #1}
              \end{center}}
        \fi}

\newcommand{\tcaption}[1]{
        \refstepcounter{table}
        \setbox\@tempboxa = \hbox{\footnotesize Table~\thetable. #1}
        \ifdim \wd\@tempboxa > 6in
           {\begin{center}
        \parbox{6in}{\footnotesize\baselineskip=12pt Table~\thetable. #1}
            \end{center}}
        \else
             {\begin{center}
             {\footnotesize Table~\thetable. #1}
              \end{center}}
        \fi}

\def\@citex[#1]#2{\if@filesw\immediate\write\@auxout
	{\string\citation{#2}}\fi
\def\@citea{}\@cite{\@for\@citeb:=#2\do
	{\@citea\def\@citea{,}\@ifundefined
	{b@\@citeb}{{\bf ?}\@warning
	{Citation `\@citeb' on page \thepage \space undefined}}
	{\csname b@\@citeb\endcsname}}}{#1}}

\newif\if@cghi
\def\cite{\@cghitrue\@ifnextchar [{\@tempswatrue
	\@citex}{\@tempswafalse\@citex[]}}
\def\citelow{\@cghifalse\@ifnextchar [{\@tempswatrue
	\@citex}{\@tempswafalse\@citex[]}}
\def\@cite#1#2{{$\null^{#1}$\if@tempswa\typeout
	{IJCGA warning: optional citation argument
	ignored: `#2'} \fi}}

 1
 1
 1

\font\ninerm=cmr9

\font\eightrm=cmr8

\begin{document}

\newcommand{\st}{\scriptstyle}
\newcommand{\sst}{\scriptscriptstyle}
\newcommand{\mco}{\multicolumn}
\newcommand{\epp}{\epsilon^{\prime}}
\newcommand{\vep}{\varepsilon}
\newcommand{\ra}{\rightarrow}
\newcommand{\ppg}{\pi^+\pi^-\gamma}
\newcommand{\vp}{{\bf p}}
\newcommand{\ko}{K^0}
\newcommand{\kb}{\bar{K^0}}
\newcommand{\al}{\alpha}
\newcommand{\ab}{\bar{\alpha}}
\def\be{\begin{equation}}
\def\ee{\end{equation}}
\def\bea{\begin{eqnarray}}
\def\eea{\end{eqnarray}}
\def\CPbar{\hbox{{\rm CP}\hskip-1.80em{/}}}

%
%
%
\def\and{{\it\&}}
\def\half{{1\over2}}
\def\third{{1\over3}}
\def\quarter{{1\over4}}
\def\rii{\sqrt{2}}
\def\to{\rightarrow}
\def\S{\mathhexbox279}
\def\gesim{\,{\raise-3pt\hbox{$\sim$}}\!\!\!\!\!{\raise2pt\hbox{$>$}}\,}
\def\lesim{\,{\raise-3pt\hbox{$\sim$}}\!\!\!\!\!{\raise2pt\hbox{$<$}}\,}
\def\boldoverdot{\,{\raise6pt\hbox{\bf.}\!\!\!\!\>}}
\def\re{{\bf Re}}
\def\im{{\bf Im}}
\def\ie{{\it i.e.}}
\def\cf{{\it cf.}\ }
\def\ibid{{\it ibid.}\ }
\def\etal{{\it et. al.}}
\def\acal{{\cal A}}
\def\bcal{{\cal B}}
\def\ccal{{\cal C}}
\def\dcal{{\cal D}}
\def\ecal{{\cal E}}
\def\fcal{{\cal F}}
\def\gcal{{\cal G}}
\def\hcal{{\cal H}}
\def\ical{{\cal I}}
\def\jcal{{\cal J}}
\def\kcal{{\cal K}}
\def\lcal{{\cal L}}
\def\mcal{{\cal M}}
\def\ncal{{\cal N}}
\def\ocal{{\cal O}}
\def\pcal{{\cal P}}
\def\qcal{{\cal Q}}
\def\rcal{{\cal R}}
\def\scal{{\cal S}}
\def\tcal{{\cal T}}
\def\ucal{{\cal U}}
\def\vcal{{\cal V}}
\def\wcal{{\cal W}}
\def\xcal{{\cal X}}
\def\ycal{{\cal Y}}
\def\zcal{{\cal Z}}
\def\aa{{\bf a}}
\def\bb{{\bf b}}
\def\cc{{\bf c}}
\def\dd{{\bf d}}
\def\ee{{\bf e}}
\def\ff{{\bf f}}
\def\ggg{{\bf g}}
\def\hh{{\bf h}}
\def\ii{{\bf i}}
\def\jj{{\bf j}}
\def\kk{{\bf k}}
\def\lll{{\bf l}}
\def\mm{{\bf m}}
\def\nn{{\bf n}}
\def\oo{{\bf o}}
\def\pp{{\bf p}}
\def\qq{{\bf q}}
\def\rr{{\bf r}}
\def\ss{{\bf s}}
\def\tt{{\bf t}}
\def\uu{{\bf u}}
\def\vv{{\bf v}}
\def\ww{{\bf w}}
\def\xx{{\bf x}}
\def\yy{{\bf y}}
\def\zz{{\bf z}}
\def\AA{{\bf A}}
\def\BB{{\bf B}}
\def\CC{{\bf C}}
\def\DD{{\bf D}}
\def\EE{{\bf E}}
\def\FF{{\bf F}}
\def\GG{{\bf G}}
\def\HH{{\bf H}}
\def\II{{\bf I}}
\def\JJ{{\bf J}}
\def\KK{{\bf K}}
\def\LL{{\bf L}}
\def\MM{{\bf M}}
\def\NN{{\bf N}}
\def\OO{{\bf O}}
\def\PP{{\bf P}}
\def\QQ{{\bf Q}}
\def\RR{{\bf R}}
\def\SS{{\bf S}}
\def\TT{{\bf T}}
\def\UU{{\bf U}}
\def\VV{{\bf V}}
\def\WW{{\bf W}}
\def\XX{{\bf X}}
\def\YY{{\bf Y}}
\def\ZZ{{\bf Z}}
\def\iBB{ \hbox{{\smallii I}}\!\hbox{{\smallii I}} }
\def\bBB{ \hbox{{\smallii I}}\!\hbox{{\smallii B}} }
\def\pBB{ \hbox{{\smallii I}}\!\hbox{{\smallii P}} }
\def\rBB{ \hbox{{\smallii I}}\!\hbox{{\smallii R}} }
\def\alpbf{{\pmb{$\alpha$}}}
\def\betbf{{\pmb{$\beta$}}}
\def\gambf{{\pmb{$\gamma$}}}
\def\delbf{{\pmb{$\delta$}}}
\def\epsbf{{\pmb{$\epsilon$}}}
\def\zetbf{{\pmb{$\zeta$}}}
\def\etabf{{\pmb{$\eta$}}}
\def\thebf{{\pmb{$\theta$}}}
\def\varthebf{{\pmb{$\vartheta$}}}
\def\iotbf{{\pmb{$\iota$}}}
\def\kapbf{{\pmb{$\kappa$}}}
\def\lambf{{\pmb{$\lambda$}}}
\def\mubf{{\pmb{$\mu$}}}
\def\nubf{{\pmb{$\nu$}}}
\def\xibf{{\pmb{$\xi$}}}
\def\pibf{{\pmb{$\pi$}}}
\def\varpibf{{\pmb{$\varpi$}}}
\def\rhobf{{\pmb{$\rho$}}}
\def\sigbf{{\pmb{$\sigma$}}}
\def\taubf{{\pmb{$\tau$}}}
\def\upsbf{{\pmb{$\upsilon$}}}
\def\phibf{{\pmb{$\phi$}}}
\def\varphibf{{\pmb{$\varphi$}}}
\def\chibf{{\pmb{$\chi$}}}
\def\psibf{{\pmb{$\psi$}}}
\def\omebf{{\pmb{$\omega$}}}
\def\Gambf{{\pmb{$\Gamma$}}}
\def\Delbf{{\pmb{$\Delta$}}}
\def\Thebf{{\pmb{$\Theta$}}}
\def\Lambf{{\pmb{$\Lambda$}}}
\def\Xibf{{\pmb{$\Xi$}}}
\def\Pibf{{\pmb{$\Pi$}}}
\def\Sigbf{{\pmb{$\sigma$}}}
\def\Upsbf{{\pmb{$\Upsilon$}}}
\def\Phibf{{\pmb{$\Phi$}}}
\def\Psibf{{\pmb{$\Psi$}}}
\def\Omebf{{\pmb{$\Omega$}}}
\def\ssb{spontaneous symmetry breaking}
\def\vev{vacuum expectation value}
\def\irrep{irreducible representation}
\def\lhs{left hand side\ }
\def\rhs{right hand side\ }
\def\Ssb{Spontaneous symmetry breaking\ }
\def\Vev{Vacuum expectation value}
\def\Irrep{Irreducible representation}
\def\Lhs{Left hand side\ }
\def\Rhs{Right hand side\ }
\def\tr{ \hbox{tr}}
\def\det{\hbox{det}}
\def\Tr{ \hbox{Tr}}
\def\Det{\hbox{Det}}
\def\diag{\hbox{\diag}}
\def\sm{Standard Model}
\def\ev{\hbox{eV}}
\def\kev{\hbox{keV}}
\def\mev{\hbox{MeV}}
\def\gev{\hbox{GeV}}
\def\tev{\hbox{TeV}}
\def\milm{\hbox{mm}}
\def\cm{\hbox{cm}}
\def\m{\hbox{m}}
\def\km{\hbox{km}}
\def\gr{\hbox{gr}}
\def\kg{\hbox{kg}}
%
%
%
\def\noteeye{ {{$\quad(\!(\subset\!\!\!\!\bullet\!\!\!\!\supset)\!)\quad$}}}
\def\note#1{{\bf \noteeye\nobreak #1 \noteeye } }
\def\doubleundertext#1{
{\undertext{\vphantom{y}#1}}\par\nobreak\vskip-\the\baselineskip\vskip4pt%
\undertext{\hbox to 2in{}}}
\def\inbox#1{\vbox{\hrule\hbox{\vrule\kern5pt
     \vbox{\kern5pt#1\kern5pt}\kern5pt\vrule}\hrule}}
\def\sqr#1#2{{\vcenter{\hrule height.#2pt
      \hbox{\vrule width.#2pt height#1pt \kern#1pt
         \vrule width.#2pt}
      \hrule height.#2pt}}}
\def\square{\mathchoice\sqr56\sqr56\sqr{2.1}3\sqr{1.5}3}
\def\today{\ifcase\month\or
  January\or February\or March\or April\or May\or June\or
  July\or August\or September\or October\or November\or December\fi
  \space\number\day, \number\year}
\def\pmb#1{\setbox0=\hbox{#1}%
  \kern-.025em\copy0\kern-\wd0
  \kern.05em\copy0\kern-\wd0
  \kern-.025em\raise.0433em\box0 }
\def\up#1{^{\left( #1 \right) }}
\def\lowti#1{_{{\rm #1 }}}
\def\inv#1{{1\over#1}}
\def\deriva#1#2#3{\left({\partial #1\over\partial #2}\right)_{#3}}
\def\su#1{{SU(#1)}}
\def\ui{U(1)}
\def\antes{}
\def\despues{.}
\def\dss{ {}^2 }
%
\def\sumprime_#1{\setbox0=\hbox{$\scriptstyle{#1}$}
  \setbox2=\hbox{$\displaystyle{\sum}$}
  \setbox4=\hbox{${}'\mathsurround=0pt$}
  \dimen0=.5\wd0 \advance\dimen0 by-.5\wd2
  \ifdim\dimen0>0pt
  \ifdim\dimen0>\wd4 \kern\wd4 \else\kern\dimen0\fi\fi
\mathop{{\sum}'}_{\kern-\wd4 #1}}
%
%
%
%
\font\sanser=cmssq8
\font\sanseru=cmssq8 scaled\magstep1 
\font\sanserd=cmssq8 scaled\magstep2 
\font\sanseri=cmssq8 scaled\magstep1
\font\sanserii=cmssq8 scaled\magstep2
\font\sanseriii=cmssq8 scaled\magstep3
\font\sanseriv=cmssq8 scaled\magstep4
\font\sanserv=cmssq8 scaled\magstep5
\font\tyt=cmtt10
\font\tyti=cmtt10 scaled\magstep1
\font\tytii=cmtt10 scaled\magstep2
\font\tytiii=cmtt10 scaled\magstep3
\font\tytiv=cmtt10 scaled\magstep4
\font\tytv=cmtt10 scaled\magstep5
\font\slanti=cmsl10 scaled\magstep1
\font\slantii=cmsl10 scaled\magstep2
\font\slantiii=cmsl10 scaled\magstep3
\font\slantiv=cmsl10 scaled\magstep4
\font\slantv=cmsl10 scaled\magstep5
\font\bigboldi=cmbx10 scaled\magstep1
\font\bigboldii=cmbx10 scaled\magstep2
\font\bigboldiii=cmbx10 scaled\magstep3
\font\bigboldiv=cmbx10 scaled\magstep4
\font\bigboldv=cmbx10 scaled\magstep5
\font\small=cmr8
\font\smalli=cmr8 scaled\magstep1
\font\smallii=cmr8 scaled\magstep2
\font\smalliii=cmr8 scaled\magstep3
\font\smalliv=cmr8 scaled\magstep4
\font\smallv=cmr8 scaled\magstep5
\font\ital=cmti10
\font\itali=cmti10 scaled\magstep1
\font\italii=cmti10 scaled\magstep2
\font\italiii=cmti10 scaled\magstep3
\font\italiv=cmti10 scaled\magstep4
\font\italv=cmti10 scaled\magstep5
\font\smallit=cmmi7
\font\smalliti=cmmi7 scaled\magstep1
\font\smallitii=cmmi7 scaled\magstep2
\font\rmi=cmr10 scaled\magstep1
\font\rmii=cmr10 scaled\magstep2
\font\rmiii=cmr10 scaled\magstep3
\font\rmiv=cmr10 scaled\magstep4
\font\rmv=cmr10 scaled\magstep5
\font\eightrm=cmr8
\def\title{}
\input epsf
\def\leff{\lcal\lowti{eff}}
\def\whatjournal{N}
\newlinechar=`\^^J
\def\orderprd#1#2#3{{\bf#1}, #2 (#3)}
\def\ordernpb#1#2#3{{\bf#1} (#3) #2}
\if P\whatjournal {\global\def\order#1#2#3{\orderprd{#1}{#2}{#3}}}
                    \immediate\write16{^^J PRD references ^^J}\else
                   {\global\def\order#1#2#3{\ordernpb{#1}{#2}{#3}}}
                    \immediate\write16{^^J NPB references ^^J}
\fi

\def\ap#1#2#3{{\it Ann. Phys.\ }\order{#1}{#2}{#3}}
\def\app#1#2#3{{\it Acta Phys. Pol. {\bf B}}\order{#1}{#2}{#3}}
\def\ijmpa#1#2#3{{\it Int. J. of Mod. Phys. {\bf A}}\order{#1}{#2}{#3}}
\def\mpla#1#2#3{{\it  Mod. Phys. Lett. {\bf A}}\order{#1}{#2}{#3}}
\def\nim#1#2#3{{\it Nucl. Instrum. \& Methods {\bf B}}\order{#1}{#2}{#3}}
\def\npb#1#2#3{{\it Nucl. Phys. {\bf B}}\order{#1}{#2}{#3}}
\def\physa#1#2#3{{\it Physica {\bf A}}\order{#1}{#2}{#3}}
\def\plb#1#2#3{{\it Phys. Lett. {\bf B}}\order{#1}{#2}{#3}}
\def\pl#1#2#3{{\it Phys. Lett.\ }\order{#1}{#2}{#3}}
\def\pr#1#2#3{{\it Phys. Rev.\ }\order{#1}{#2}{#3}}
\def\prep#1#2#3{{\it Phys. Rep.\ }\order{#1}{#2}{#3}}
\def\prl#1#2#3{{\it Phys. Rev. Lett.\ }\order{#1}{#2}{#3}}
\def\prd#1#2#3{{\it Phys. Rev. {\bf D}}\order{#1}{#2}{#3}}
\def\ptp#1#2#3{{\it Prog. Theo. Phys.\ }\order{#1}{#2}{#3}}
\def\rmp#1#2#3{{\it Rev. Mod. Phys.\ }\order{#1}{#2}{#3}}
\def\zphys#1#2#3{{\it Z. Phys. {\bf C}}\order{#1}{#2}{#3}}
\centerline{\normalsize\bf APPLICATIONS OF EFFECTIVE
LAGRANGIANS\footnote{Talk given at {\sl Beyond the Standard Model IV},
 Lake Tahoe, CA, Dec. 13-18, 1994}}
\vspace*{0.6cm}
\centerline{\footnotesize JOS\'E WUDKA}
\baselineskip=13pt
\centerline{\footnotesize\it Physics Department, University of California,
Riverside}
\baselineskip=12pt
\centerline{\footnotesize\it Riverside, CA 92521-0413, U.S.A.}
\centerline{\footnotesize E-mail: wudka@ucrph0.ucr.edu}

\vspace*{0.9cm}
\abstracts{ 
The applications of effective lagrangians to the determination of the
effects of physics beyond the \sm\ are briefly described. Emphasis is
given to those effective operators which generate the
largest deviations form the \sm; some applications are described.
}

\normalsize\baselineskip=15pt
\setcounter{footnote}{0}
\renewcommand{\thefootnote}{\alph{footnote}}
\section{Introduction}
The \sm\ is  a theory in perfect agreement with all experimental data.
Its predictive power is very large and is accepted as a correct
description of nature at all scales which have been probed (up to
energies $ \sim 100 \gev $). Nonetheless the model has
various theoretical problems (such as the possible triviality of the
scalar sector~\cite{Callaway}) and it is the consensus of the community
that it represents the low energy limit of a more fundamental theory.

There have been many proposed extensions of the \sm, mainly based on
specific kinds of new physics imagined to be apparent at
energies significantly above the ones currently probed. Typical examples
of these approaches are the supersymmetric~\cite{SUSY} and
technicolor~\cite{TC} pictures of
new physics. Should a future experiments discover a techni-rho, or a
slepton or wino, the guessing game would be over and all efforts
will be concentrated in elucidating the specific technicolor or
supersymmetric model realized in nature.

There is, however, the possibility that we will not be able to
directly observe the effects of the new dynamics
(via the creation of new particles); then only virtual effects are
available as probes into the
physics beyond the \sm. For this situations the most concise approach
available is based on the use of effective lagrangians~\cite{Wein/Geo,Pol}.

Suppose for the moment that the lagrangian $ \lcal\lowti{new} $, which
describes the physics beyond the \sm, is known. Suppose that
non-standard effects become directly observable at center of mass
energies of the order of a scale $ \Lambda $. If we wish to obtain a
description of the non-standard effects generated by $ \lcal\lowti{new} $
at low energies, one should integrate out all heavy fields (of mass $ \sim
\Lambda $ or higher) and determine the corresponding effective action
(which, by construction, contains only \sm\ fields). This effective
action will also contain the scale $ \Lambda $ as a parameter.

Since we are interested in physics at energies significantly below $
\Lambda $,  a large-$ \Lambda $ expansion is appropriate. The result
of performing this expansion on the effective action obtained above is a
tower of {\it gauge-invariant} local operators (which I denote by $ \ocal_i $)
containing the \sm\ fields multiplied by some
function of $ \Lambda $ and the couplings present in $ \lcal\lowti{new}
$, namely
\begin{equation}
 S\lowti{eff} = \int d^4 x \leff ; \quad
\leff = \sum_i f_i ( \Lambda ; \hbox{couplings} ) \ocal_i  . \end{equation}
The dependence on  $ \Lambda $ of each term is determined by the
dimension of the operator in  question (up to factors of $ \ln \Lambda
$): $f_i \propto \Lambda^{ 4 - \dim \ocal_i } $. Thus I can write
\begin{equation}
\leff = \Lambda^4 \sum_i { \alpha_i ( \hbox{ couplings} )
\over \Lambda^{ d_i } } \ocal_i ; \qquad d_i = \hbox{ dim } \ocal_i
\end{equation}
All operators will be generated (in general)
irrespective of the specific details of $ \lcal\lowti{new} $.
In contrast the $ \alpha_i $ are model specific and summarize
all the information that can be gathered from $ \lcal\lowti{new} $ at
low energies.
One can then use these parameters to parametrize all new physics effects
without model prejudices~\cite{Wudka}.

\section{Symmetries of the effective lagrangian}
It was stated above that the local symmetries of the \sm\ are
preserved by the effective lagrangian. An effective lagrangian which
does not satify this condition presents severe self-consistency
problems~\cite{Veltman}. Suppose for example
that a gauge-variant three-gauge-boson vertex is introduced into an
otherwise gauge invariant model. We can the  consider the corrections to
the vector boson masses generated by this term

\setbox2=\vbox to 25 pt{\epsfxsize=4in\epsfbox[0 0 612 622]{f1_bsmiv.ps}}

\begin{equation} \box2  \mskip-150mu
:\quad \delta m^2 \sim m^2 \left( { g_H \over 4 \pi } { \Lambda^3 \over m^3 }
\right)^2 \end{equation} where $ m$ denotes the vector-boson-mass, $ \Lambda $
is
the scale of the physics generating the gauge-variant term and $g_H$ the
corresponding coupling constant. For the \sm\ the corrections to $m$ must be
small (due to the agreement with the data) hence $ \Lambda \ll m ( 4 \pi
/ g_H )^{1/3} $. For $ m_W $ we know that $ \delta m_W^2 / m_W^2 \lesim
0.0064 $ which implies $ \Lambda \lesim m_W /2 $. If no restrictions are
placed on $ \Lambda $ radiative corrections shift $m \rightarrow O (
\Lambda ) $.

One might wonder whether it is only the global symmetries associated
with the gauge invariance that should be imposed on $ \leff $. But in
this case there is no cogent reason for lepton universality; moreover,
there will be no connection between the cubic and quartic vector-boson
couplings and the above problems with the masses reappear~\footnote{
Of course one can decide to forego gauge invariance completely and fine
tune everything. This is possible but quite useless since such a theory
has absolutely no predictive power.}

These arguments do not apply to global symmetries.
Consider for example a term in $ \leff $ of the form
\begin{equation}
 \inv\Lambda \alpha_{ e \mu B
} \bar e_R \sigma_{ \mu \nu } \mu_R B^{ \mu \nu } \end{equation} where $ e_R ,
\mu_R $
denote the corresponding right-handed fermion fields and $B_{ \mu \nu } $
is the field-strength for the $ \ui_Y $ gauge field. This interaction
generates a non-vanishing branching ratio for the process $ \mu
\rightarrow e \gamma $ proportional to $ \left| \alpha_{ e \mu B } /
\Lambda \right|^2 $; the fact that this transition highly forbidden
merely indicates that $ \Lambda $ is very large.

\section{Low energy particle content}
The recipe for constructing the effective lagrangian is to select the
light excitations to be studied and the symmetries that they are to
respect and then to construct the most general set of local operators
involving the corresponding fields and respecting the said symmetries.

For the \sm\ we have the choice of including a scalar sector as low
energy excitations or not. If there is a light Higgs and I assume that $
\Lambda \not= v $ (where $v$ is the \sm\ \vev) then the decoupling
theorem~\cite{Dec. thm} insures that all observables appear as a power series
in $
1 / \Lambda $, in particular, all effects from the new physics disappear
as $ \Lambda \rightarrow \infty $~\footnote{ Note that the condition $
\Lambda \not= v $ excludes, for example, a heavy fourth generation.}.

If there be no light Higgs then the imsplest way of descibing the
scalar sector (we still need the Goldstone bosons in order
generate masses for the $W$ and
$Z$) is through a non-linear sigma model
{}~\cite{Coleman,Wein/Geo}. In this case the decoupling theorem
is not applicable. Due to time constraints I will
not discuss this situation in this talk.

Consider therefore the situation where there is a light Higgs present.
Then the effective lagrangian takes the form
\begin{equation}
 \leff = \lcal_{ SM }
+ \sum_i { \tilde \alpha_i \over \Lambda } \ocal \up{\rm dim\ 5 }_i
+ \sum_i { \alpha_i \over \Lambda } \ocal \up{\rm dim\ 6 }_i + \cdots .
\end{equation}
It is easy to see that there
are no dimension 5 operators satisfying the \sm\ symmetries;
there are 81 dimension-six operators~\cite{BW} (for one family of
fermions).

For the case of a light Higgs I'll require that it's mass not be
shifted to $ O ( \Lambda ) $ by radiative corrections; it follows that
the underlying theory
should be assumed to be weakly coupled. We can avoid this
constraint by either fine-tuning or by modifying the low-energy
particle content as in supersymmetry. I will not consider either of
these situations, the first is unnatural and the second requries a
low-energy sector substantially different from the on in the \sm. As
indicated in the introduction I assume no direct
observation of non-\sm\ physics is available; this excludes the second case.
\section{Tree-level operators.}
For the above scenario the relevant
property of a given $ \ocal_i $ is whether it is tree-level or loop
generated by the underlying dynamics. Loop generated operators appear with a
characteristic suppression factor $ \sim 1 / ( 4 \pi )^2 $ which
significantly decreases the magnitude of their effects. Assuming that the
underlying theory is a gauge theory, the only tree level generated
operators take the form~\cite{Arzt}
\begin{equation}
\left( \phi D \phi \right)^2 , \qquad
\phi^6 , \qquad
\left( \phi \psi ) D \right( \phi \psi ) , \qquad
\phi^3 \psi^2  \qquad
\psi^4 , \end{equation} where $ \phi $ denotes the \sm\ scalar doublet and $
\psi $ a light fermion; $D$ denotes the covariant derivative. The
notation used above is, of course symbolical.
%
\medskip
{\footnotesize
To see how the list of operators was obtained~\cite{Arzt}
consider a broken gauge
and let $V$ be the corresponding \vev. At this level the \sm\ subgroup
will be left unbroken. The
unbroken subgroup will be $ \su3 \times \su2 \times \ui $ and the
corresponding (light) generators will be denoted generically by $ T^\ell $;
broken (or heavy) generators will be labelled $ T^h $; it follows
that $ T^\ell V = 0 $ while $ T^h V \not= 0 $. I will denote by $X$ the
heavy vector bosons and by $W$ the light ones; light and heavy scalars
are denoted by $ \phi $ and $ \Phi $ respectively (fermions can be treated
in the same way). The structure constants
of the full theory will be denoted by $ f $.

\begin{itemize}
\item $ X W W $ vertices will be proportional to $ f_{
\ell \ell h } $. But, since the unbroken generators form a group, the
commutator of two of them should also give a light generator; hence
$ f_{ \ell \ell h } = 0 $ and that this type of vertices is absent.

\item $ X W \Phi $ vertices will be proportional to $ V
T^h T^\ell \Phi $; the vector $ V T^h $ lies along a Goldstone boson
direction. Since the light group is unbroken,
the action on a Goldstone boson direction by $ T^\ell $ must also
give a Goldstone boson direction (since the Goldstone bosons transform as
a representation of the unbroken group). But then $ V T^h T^\ell $ will
be orthogonal to all directions corresponding to heavy scalars: this
type of vertices is absent also.
\end{itemize}

Once all the vertices of the heavy theory have been studied along these
lines it is straightforward to determine, given all the possible
tree-level graphs that can be drawn (and which correspond to
dimension-six operators), which are actually generated by a gauge
theory. }
\medskip

For the \sm\ the tree-level generated dimension six operators are
$$
\ocal_\phi = \inv3 \left( \phi^\dagger \phi \right)^3
\qquad \ocal_{ \partial \phi } = \half \left[ \partial \left( \phi^\dagger \phi
\right) \right] ^2
\qquad \ocal\up1_\phi = \left( \phi^\dagger \phi \right) \left| D_\mu \phi
\right|^2
\qquad \ocal\up3_\phi = \left| \phi^\dagger D \phi \right|^2
$$
$$
\ocal\up1_{ \phi \ell } = i \left(\phi^\dagger D_\mu \phi \right) \bar\ell
\gamma^\mu \ell
\qquad \ocal\up3_{ \phi \ell } = i \left(\phi^\dagger \tau^I D_\mu \phi \right)
\bar\ell \tau^I \gamma^\mu \ell
\qquad \ocal_{ \phi e } = i \left( \phi^\dagger D_\mu \phi \right) \bar e
\gamma^\mu e
$$
$$
\ocal\up1_{ \phi q } = i \left(\phi^\dagger D_\mu \phi \right) \bar q
\gamma^\mu q
\qquad \ocal\up3_{ \phi q } = i \left(\phi^\dagger \tau^I D_\mu \phi \right)
\bar q \tau^I \gamma^\mu q
$$
\begin{equation}
\ocal_{ \phi u } = i \left( \phi^\dagger D_\mu \phi \right) \bar u \gamma^\mu u
\qquad \ocal_{ \phi d } = i \left( \phi^\dagger D_\mu \phi \right) \bar d
\gamma^\mu d
\qquad \ocal_{ \phi  \phi } = i \left( \phi^T \epsilon D_\mu \phi \right)
\left( \bar u \gamma^\mu d \right)
\end{equation}
where
$ \phi = $\sm\ scalar doublet,
$ D = $covariant derivative,
$ \ell = $left-handed lepton doublet,
$ e = $right-handed (charged) lepton singlet,
$ q = $left-handed quark doublet,
$ u = $right-handed u-quark singlet,
$ d = $right-handed d-quark singlet, and
$ \epsilon = i \; \tau^2 $.

Note that this list does not contain any terms that modifies the $WWZ$
and $WW\gamma $ couplings. This implies that all such modification will
be loop generated. I  terms of the now standard notation for the
parameters describing the anomalous $WWZ$ and $WW\gamma $ vertices this
implies~\cite{Wudka}
\begin{equation}
\left| \lambda \right| \sim { 10^{ - 4 } \over \Lambda \lowti{ TeV } }
\qquad
\left| \kappa - 1 \right| \sim { 2 \times 10^{ - 4 } \over \Lambda \lowti{ TeV
} }
\end{equation}
where $ \Lambda \lowti{ TeV } $ denotes  $ \Lambda $ in \tev\
units. This implies that a bound $ | \lambda | \lesim 1 $ implies $
\Lambda \gesim 10 \gev $: for loop generated operators most of the
current data lack the precision necessary to probe interesting regions
of the $ \Lambda $ axis. The fact that there are no deviations from the
\sm\ in this area is far from surprising.


\section{Some bounds from current data.}

The tree level operators considered above have various effects
on various observables which have been measured to high precision
at LEP1,  namely, the couplings of the leptons to the $Z$ boson.
Using the lagrangian $ \leff = \lcal_{ SM } +
\sum_i \left( \alpha_i \ocal_i + \hbox{ h.c. } \right) / \Lambda^2 $
the vector and axial couplings of the electron to the $Z$ and the
neutrino couplings to the $Z$ are modified according to
  \begin{eqnarray}
\left| \delta g_V ( e ) \right| &=& { v^2 \over \Lambda^2 } \left| \alpha_{
\phi \ell }\up1 + \alpha_{ \phi \ell }\up3 + \alpha_{ \phi e } \right| \lesim
0.0021 \\ \nonumber
\left| \delta g_A ( e ) \right| &=& { v^2 \over \Lambda^2 } \left| \alpha_{
\phi \ell }\up1 + \alpha_{ \phi \ell }\up3 - \alpha_{ \phi e } \right| \lesim
0.00064 \\ \nonumber
\left| \delta  g( \nu ) \right| &=& { v^2 \over \Lambda^2 } \left| \alpha_{
\phi \ell }\up1 - \alpha_{ \phi \ell }\up3 \right| \lesim 0.0018 \\ \nonumber
\end{eqnarray}
where the bounds correspond to the $ 1 \sigma $ errors given in
the particle data book~\cite{pdb}.
The $ 3 \sigma $
limits obtained from these results are~\footnote{These values
are somewhat different
from the ones presented at the talk due to an algebraic error;
these are the correct results.}
  \begin{equation}
 \Lambda\lowti{ TeV }
\gesim 2.5 \sqrt{ \alpha_{ \phi \ell }\up1 }
, \quad 2.5 \sqrt{ \alpha_{ \phi \ell }\up3 }
, \quad 2.7 \sqrt{ \alpha_{ \phi e } } \end{equation}

These bounds obtained above are quite significant; they
preempt many new results that could be obtained from LEP2 when considering
the couplings of fermions to the gauge bosons.

Just as for the couplings of the leptons to the $Z$ one can easily
derive the modifications to the oblique parameters~\cite{Pesk/Tak}
generated by
the effective operators of dimension six. The expressions for these
modifications and the contributing operators are~\cite{Wudka}
\begin{equation}
\begin{array}{l @{\qquad} l}
\delta S = 32 \pi \alpha_{ W B } ( v^ 2 / \Lambda^2 ) &
\ocal_{ W B } = \left( \phi^\dagger \tau^I \phi \right) W_{ \mu \nu  }^I B^{
\mu \nu } \hfill
\\
\delta T = - { 4 \pi \over s^2\lowti w } \alpha_\phi\up3 ( v^2 / \Lambda^2 ) &
\ocal_\phi\up3 = \left| \phi^\dagger D \phi \right|^2 \hfill \\
\delta U = O ( v^4 / \Lambda^4 ) & \\
\end{array}
\end{equation} where the natural sizes
of the coefficients are $ \alpha_{ W B } \sim 1/ ( 4 \pi )^2 $ and
$ \alpha_\phi\up3 \sim 1 $.

The existing bounds on these quantities imply non-trivial
sensitivity to $ \Lambda $, namely
  \begin{equation}
\left| \delta S \right| < 0.4 \ \Rightarrow \ \Lambda \gesim 310 \gev; \quad
\left| \delta T \right| < 0.4 \ \Rightarrow \ \Lambda \gesim 2.9 \tev; \quad
\left| \delta U \right| < 1.3 \ \Rightarrow \ \Lambda \gesim 430 \gev. \quad
\end{equation}

For  the case where there is non light Higgs the natural size for the
contributions to $U$ are $ \sim 1 / \pi $. Should the uncertainty in
the $W$ mass reach $ \sim 40 \mev $ the corresponding sensitivity to
$ \Lambda $ reaches $ \sim 690 \gev $ and this measurement can be used
to differentiate, albeit indirectly, between the light Higgs and the
no-Higgs scenarios.

\section{Higgs reactions as probes of new physics.}
Most of the measurements of $ \Lambda $ to be performed at LEP2 become
redundant. It therefore becomes interesting to isolate those processes
for which the LEP2 measurements will provide new insights into
the physics beyond the \sm; the same is true for the proposed New Linear
Collider (NLC). I will concentrate the two
processes
\begin{equation}
 \hfill \bullet \; e^+ e^- \rightarrow \nu \bar \nu H \qquad \qquad
 \bullet \; e^+ e^- \rightarrow Z H . \hfill
\end{equation}
and use them as probes for possible deviations form the \sm\  generated by new
physics.

The amplitudes for these processes
take the (symbolic) form
$
 \acal = \acal_{ SM } + [ \alpha \acal_1
+ \alpha'(s / v^2) \acal_2 ] ( v^2 / \Lambda^2 )
$; obviously this type of expression cannot be used for arbitrary
values of $s$ (in fact, we already know that the parametrization  used
will certainly break down when $ s \sim \Lambda^2 $). We will only consider
values
of $s$ for which the new contributions proportional to $ \alpha, \; \alpha' $
are smaller than the \sm\ contributions.

The relevant operators are
$
\ocal_\phi, \
\ocal_{ \partial \phi }, \
\ocal\up1_\phi, \
\ocal\up3_\phi, \
\ocal\up1_{ \phi \ell }, \
\ocal\up3_{ \phi \ell } $ and $\ocal_{ \phi e }$.
Using this list and the above definition of the effective lagrangian the
cross sections can be easily derived.
The results of this calculation are presented in the figures below
(imposing the LEP1 constraints on the $ \alpha_i $) and
show that the reaction $ e^+ e^- \rightarrow \nu \bar \nu H $
cannot be used as a probe for new physics. This is not the case for
$ e^+ e^- \rightarrow Z H $.

To estimate the sensitivity of LEP2 and
NLC to new physics we evaluate the statistical significance of the
deviations from the \sm: consider the total number of events $N$
and the corresponding \sm\ prediction $ N_{ SM } $, then the
statistical significance is
$
 \ncal_{sd} = { | N - N_{ SM } | / \sqrt N }
$.
When evaluating $ \ncal_{ sd } $ we impose the restrictions generated
by LEP1, namely, given a value of $ \Lambda $ we require that the
$ \alpha_i $ satisfy the bounds on $ \delta g_{V , A } (e ) $ and
$ \delta g ( \nu ) $ (at the $ 3 \sigma $ level) as given above.
In this guise the regions above the curves in the following plot
will be inaccessible to LEP2; these curves give the maximum sensitivity
for a given $ \Lambda $ that can be obtained at LEP2 with the constraints
from LEP1 imposed.

\setbox2=\vbox to 245pt{\epsfxsize=4 in\epsfbox[0 0 612 660]{f2_bsmiv.ps}}
\centerline{ \box2  }

\noindent

\setbox2=\vbox to 160pt {\epsfxsize=200pt\epsfbox[0 0 612 682]{f3_bsmiv.ps}}
\centerline{ \box2  }

As can be seen from this plot non-trivial sensitivity to $ \Lambda $
can be attained at LEP2 and, more dramatically, at NLC.

\section{Conclusions}
\hfuzz=2pt
I have argued that the effects from the physics beyond the \sm\
can be naturally parametrized using an effective lagrangian. For
the case where there is a light Higgs naturality requires the
underlying theory to  be weakly coupled (alternatively the low
energy particle content should be significantly modified) in which
case the most sensitive probes into the new interactions correspond
to those operators that can be generated at tree level by the
new dynamics. For these operators existing LEP1 data imply
$ \Lambda / \sqrt{ \alpha } \gesim $ a few \tev.

I have also
showed that there are other reactions involving the Higgs boson,
that can generate new windows into the physics beyond the \sm.
These reactions cannot be probed using LEP1 data, but may very well
be available at LEP2, provided the Higgs is sufficiently light.

It is important to measure all coefficients of tree level operators
as the deviations form the \sm\ can be more significant in one of them
than in the rest. But even if no deviations from the \sm\ are
observed, their absence will point to the suppression of a large
class of interactions (in  the underlying theory!) and will
restrict the characterisitics of the corresponding models.

\section{Acknowledgements}
Most of the work presented here was developed in collaboration with
C. Arzt, M. Einhorn and with B. Grzadkowski.

\medskip

\end{document}